\documentclass[aps,prl,amsmath,amssymb,showpacs,twocolumn]{revtex4}

\usepackage{bbold}
\usepackage{graphicx}
\usepackage{amsmath}
\usepackage{amssymb,amsthm}

\newcommand{\ket}[1]{|#1\rangle}
\newcommand{\braket}[2]{\langle{#1}|{#2}\rangle}

\newcommand{\bra}[1]{\langle#1|}

\usepackage{hyperref}
\usepackage{color}

\def\eq{\begin{eqnarray}}
\def\en{\end{eqnarray}}

\usepackage[normalem]{ulem} 
\usepackage{ragged2e}

\usepackage[caption=false,position=t,singlelinecheck=false,justification=raggedright]{subfig}

\begin{document}

\title{
Many-body state engineering using measurements and fixed unitary dynamics
}
\author{Mads Kock Pedersen, Jens Jakob W. H. S\o rensen, Malte C. Tichy, Jacob F. Sherson}
\email{sherson@phys.au.dk}
\affiliation{AU Ideas Center for Community Driven Research, CODER, \\
 Department of Physics and Astronomy, University of Aarhus, DK--8000 Aarhus C, Denmark}

\date{\today}

\begin{abstract}
We develop a scheme to prepare a desired state or subspace in high-dimensional 
Hilbert-spaces using repeated applications of a single static projection operator 
onto the desired target and fixed unitary dynamics. Benchmarks against other 
control schemes, performed on generic Hamiltonians and on Bose?Hubbard systems, 
establish the competitiveness of the method. As a concrete application of the 
control of mesoscopic atomic samples in optical lattices we demonstrate the near 
deterministic preparation of Schrödinger cat states of all atoms residing on 
either the odd or the even sites.
\end{abstract}
\pacs{
05.30.Jp,    
03.75.Lm, 
03.75.Gg 
}
\maketitle

\begin{figure}[tbp]
\includegraphics[trim=0 10 0 10,width=.995\columnwidth]{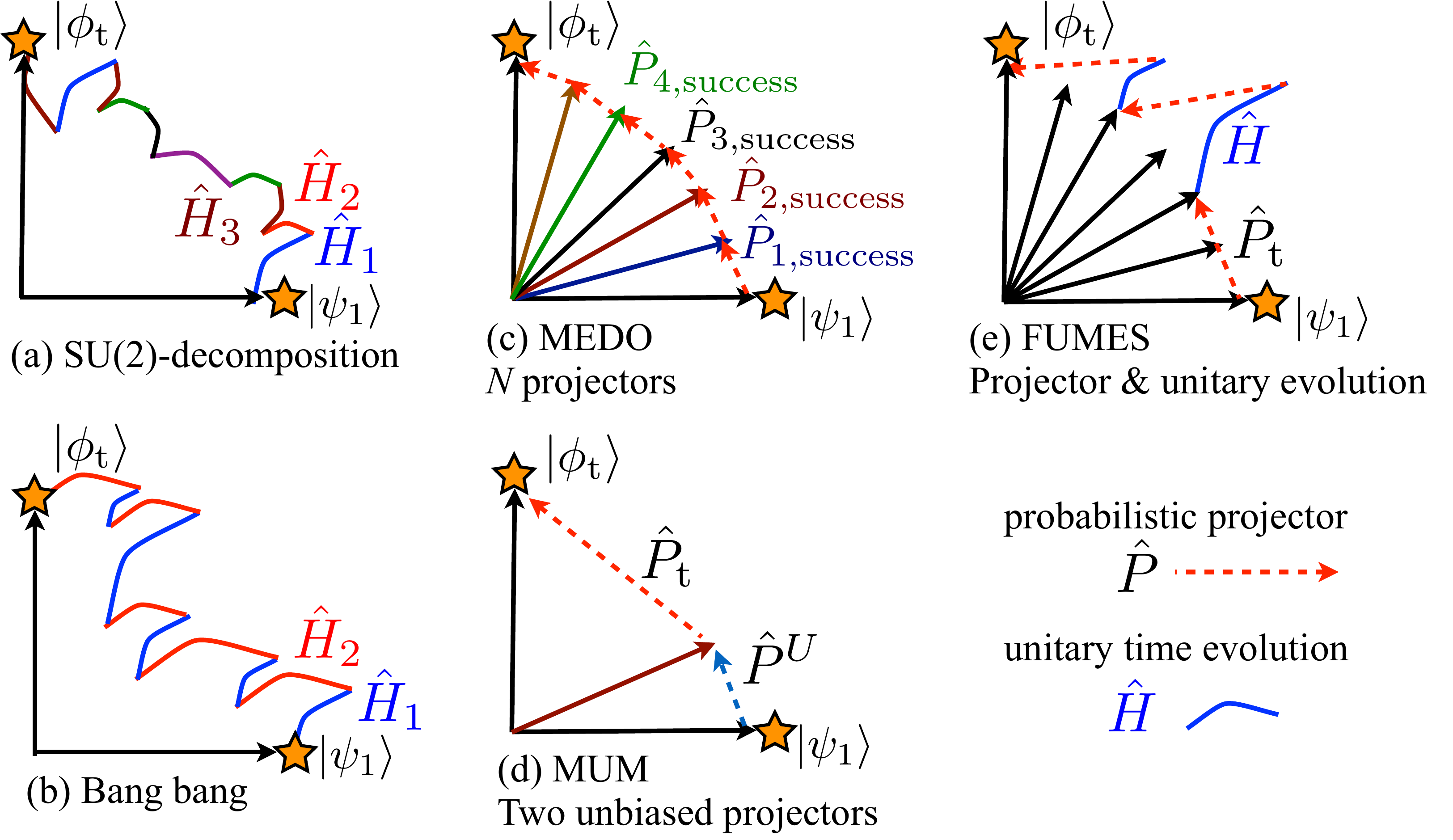}
\caption{(color online) Paradigms for quantum state engineering. (a)~SU(2)-decomposition of the unitary evolution into $d(d-1)/2$ two-level unitaries. (b)~Bang-Bang control via two non-commuting Hamiltonians. (c)~MEDO: multiple measurements successively bring the state closer to the target. (d)~MUM: two mutually unbiased observables are measured alternately. (e)~FUMES alternates the unitary evolution with measurements at optimised moments in time.}
\label{fig:Cartoon}
\end{figure}

\emph{Introduction --} Quantum state engineering aims at preparing a desired target quantum state \cite{Brif:2010ad}, which is a prerequisite for e.g. control of qudits \cite{Norris}, and for quantum technologies such as quantum computation, quantum metrology and the synthesisation and simulation of novel phases of matter \cite{Lewenstein}. When the system Hamiltonian is sufficiently controllable, tailored control pulses can produce every state via \emph{unitary control} \cite{DAlessandro2008}. Paradigmatic proofs of controllability rely on the decomposition of the desired unitary transformation into two-level unitaries \cite{PhysRevLett.73.58} [Fig.~\ref{fig:Cartoon}(a)], or on two sufficiently non-commuting Hamiltonians that are switched in a ``bang-bang'' fashion \cite{Viola1998} [Fig.~\ref{fig:Cartoon}(b)]. Although tailored applications of optimal control to many-body systems such as ultacold atoms in optical lattices have lately been applied to specific applications such as the superfluid to Mott-insulator phase-transition  via control of the lattice depth \cite{PhysRevLett.106.190501,Rosi2013}, the implementation of generic approaches to full controllability remains impractical \cite{Tichy2013}. This is true even for the moderately small precisely prepared systems of 5-10 atoms, which can now be prepared with high purity \cite{Weitenberg}.

Besides unitary control via the system Hamiltonian, quantum state engineering can also be exerted for a fixed Hamiltonian  via the back-action induced by measurements \cite{1367-2630-12-4-043005,Blok2014,roch,PhysRevA.74.052102,Ashhab2010,PhysRevA.86.062331,1742-6596-84-1-012017,Burgarth2014}. On the one hand, measurement operators that are adapted on previous measurement outcomes can steer  a quantum state into a target state \cite{1367-2630-12-4-043005,Blok2014,roch}. For example, spin correlations \cite{Hauke2013} can be induced by a sequence of standing wave probes \cite{Eckert2007}. The implementation of measurements for such adaptive schemes is, however, impractical for high-dimensional systems with natural experimental restrictions. 

On the other hand, given a projector onto the desired state $\hat P_{\text{t}}=\ket{\psi_{\text{t}}}\bra{\psi_{\text{t}}}$, an 
 initial state $\ket{\psi_1}$ can be steered into the target:  \emph{Control-free control} \cite{1742-6596-84-1-012017} uses \emph{multiple evenly distributed observables} (MEDO), i.e.~the system is projected step-wise onto states that are evenly distributed between the initial and the target state [Fig.~\ref{fig:Cartoon}(c)]. Alternatively, if -- besides the projector onto the target state $\hat P_\text{t}=\ket{\psi_{\text{t}}}\bra{\psi_{\text{t}}}$ --  only one auxiliary projector $\hat P^U$ can be used, it is advisable to choose $\hat P^U$ to be mutually unbiased with respect to $\hat P_{\text{t}}$  to maximise the probability to find the target state, i.e.~to perform \emph{mutually unbiased measurements} (MUM) [Fig.~\ref{fig:Cartoon}(d)] \cite{PhysRevA.73.012322}. These schemes assume no unitary evolution between subsequent measurements. 

\begin{figure*}[tbp]
\captionsetup[subfigure]{labelformat=empty}
\subfloat[]{\label{fig:RandHamCumProps}
\includegraphics[trim=13 60 15 70,clip,width=\columnwidth]{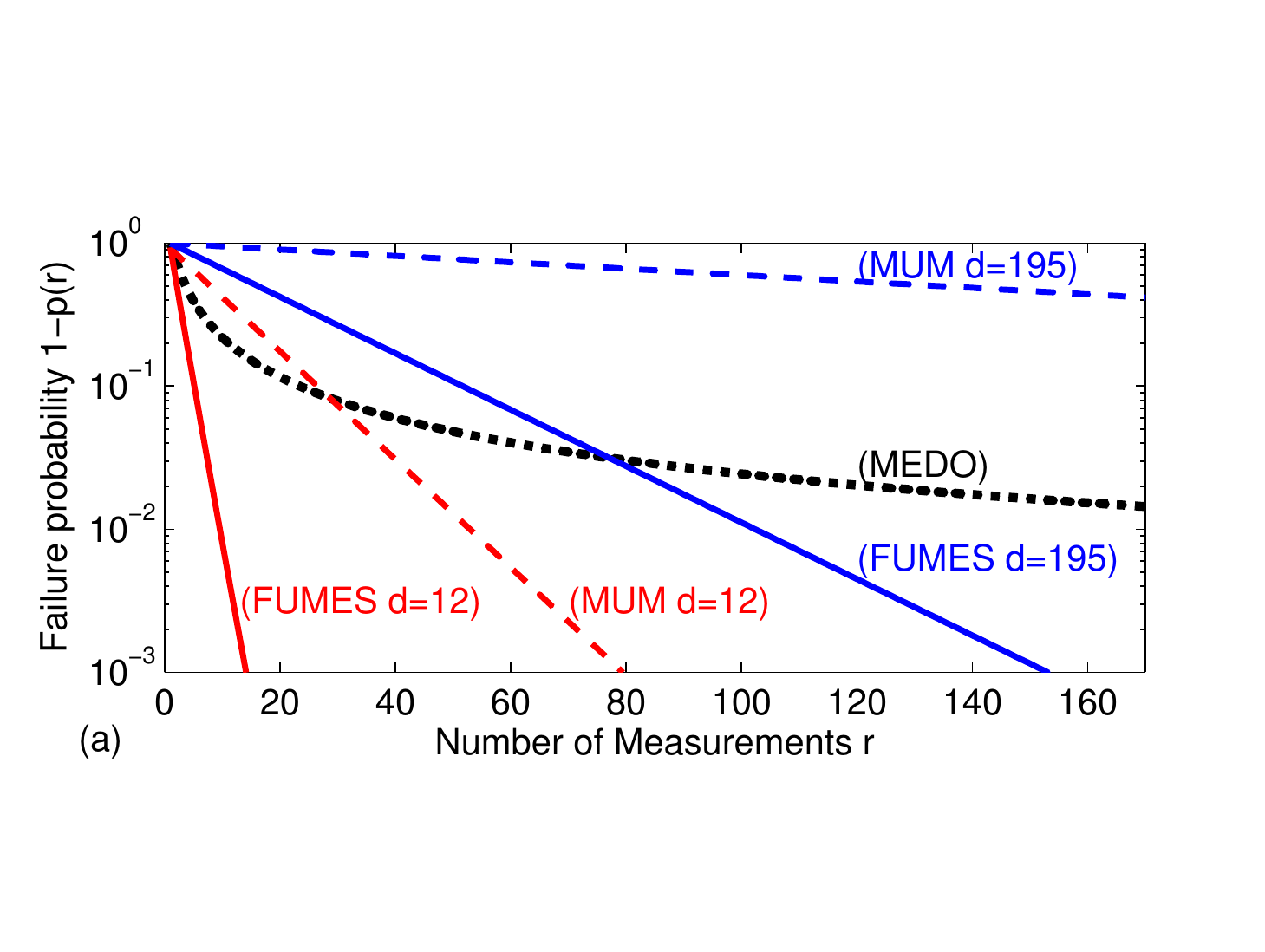}
}
\subfloat[]{\label{fig:EffectiveDimension}
\includegraphics[trim=13 60 15 70,clip,width=\columnwidth]{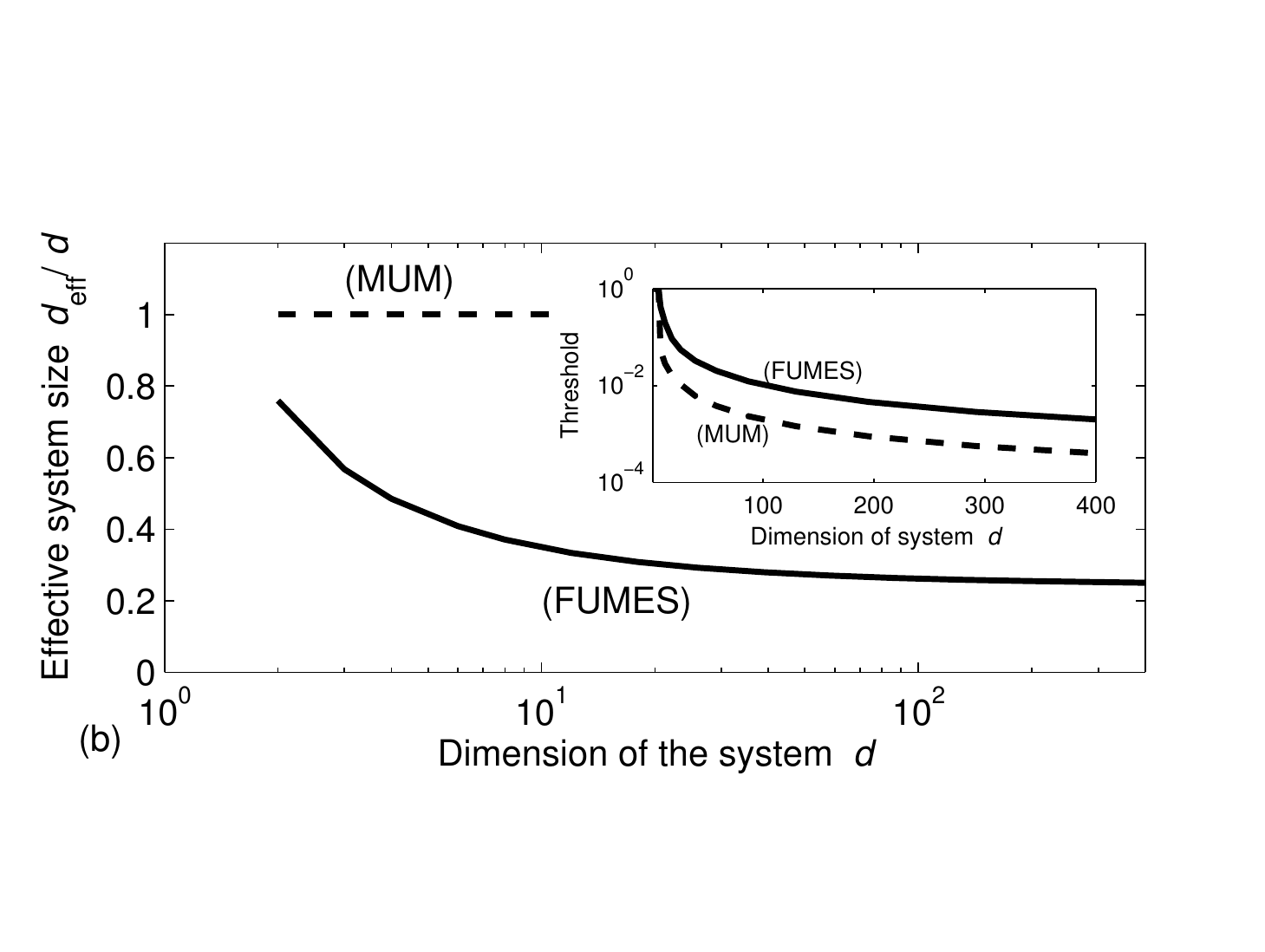}
}
\caption{(color online) (a) Probability for not reaching a randomly chosen target state after $r$ measurements.  MEDO (black dotted line) performs independently of dimensionality. Solid lines depict the average trajectory of 100,000 random Hamiltonians with FUMES for dimension $d=12$ (red) and $d=195$ (blue); dashed lines denote MUM for $d=12$ (red) and $d=195$ (blue). (b) Average normalised effective system size (\ref{eq:effectiveDim}) for MUM (dashed) and FUMES (solid) from 100,000 random Hamiltonians. (inset) Threshold failure probability below which MUM (dashed) and FUMES (solid) outperform MEDO. } 
\label{fig:RandHam}
\end{figure*}

Here, we propose a hybrid of unitary and control-free control by investigating quantum state engineering via 
optimised \emph{fixed unitary evolution and measurements} (FUMES), which imposes fairly weak requirements on the experimental infrastructure: The system Hamiltonian $\hat H$ of finite dimension $d$ is fixed and uncontrollable. The only other means to steer the system is provided by a unique fixed measurement operator. The measurement operator signals whether the target subspace is reached (eigenstate of $\hat P_{\text{t}}$, ``success''), or into which other subspace the state was projected (``failure''). A crucial feature of this protocol is the ability to model precisely the unitary many-body evolution and to determine the precise times that maximise the success probability of the subsequent measurement.  Even provided with  these minimal tools, desired states can be produced under mild assumptions on the measurement operator and the system Hamiltonian.   To benchmark the procedure, FUMES applied with a randomly chosen Hamiltonian is first compared to MUM and MEDO. Further below, we apply it to Bose-Hubbard multi-mode systems, which also permit benchmarking against unitary bang-bang control. Finally, as a practical example of FUMES, we demonstrate the nearly deterministic preparation of macroscopic superposition states.  

\emph{Quantum state engineering by FUMES --}  We denote the $d$ eigenstates of the constant Hamiltonian $H$ by  $\ket{\psi_1}, \dots, \ket{\psi_d}$. The goal is to prepare an eigenstate of a given $K$-dimensional projector $\hat P_{\text{t}}=\sum_{k=1}^K \ket{\phi_k} \bra{\phi_k}$, where $\braket{\phi_{k}}{\phi_j}=\delta_{j,k}$. For $K=1$, the target is a unique  state $\ket{\phi_{\text{t}}}$, for $K=d-1$, the aim is to prepare any state orthogonal to some $\ket{\phi^\perp}$. The system is initially in the ground state, $\ket{\psi_1}$. We mainly focus on the most stringent goal, a well-defined pure state ($K=1$), we will loosen this requirement below in our discussion of the synthesization Schr\"o{}dinger-cat-states.

The application of the projector at a random moment in time has low success probability \cite{burgarthDiss}. We therefore optimise numerically the moment in time at which a measurement is performed, maximising the probability to populate a desired eigenstate of $\hat P_{\text{t}}$. A measurement that does not yield the desired outcome described by $\hat P_{\text{t}}$ has \emph{failed} and projects the state into one of the $M$ different possible outcomes that indicate failure ($1 \le M \le d-K$). The procedure is repeated until successful, i.e.~for each failed measurement, a new optimal waiting time is chosen before the next measurement is applied.

For a \emph{binary} measurement ($M=1$), failure leads to state collapse onto an eigenstate of the $d-K$-dimensional projector $\hat Q=\mathbb{1}-\hat P_{\text{t}}$, and only marginal information is gained. A fully \emph{granular} measurement with $M=d-K$ reveals the precise state $\ket{\eta_m}$ that the system collapsed onto. 

In general, a target can be reached by FUMES if and only if no conserved quantity can be constructed out of the failed projectors $\hat Q_j$, i.e.~every sum of any $m$ projectors (${1 \le m \le M}$) does not commute with $\hat H$,
\eq 
[ \hat Q_{j_1} + \hat Q_{j_2} + ... + \hat  Q_{j_m}, \hat H ] \neq 0 , \label{sumprojectors}
\en
for any set of $\{j_l...j_k\}$.  In this case, every state that the system can be projected on evolves into the target subspace. 
Alternatively, if a combination of the $\hat Q_j$ commutes with $\hat H$, states in the corresponding eigenspace will never evolve into the target subspace. 

\begin{figure*}[tbp]
\captionsetup[subfigure]{labelformat=empty}
\subfloat[]{
\includegraphics[trim=0 0 0 0, width=\columnwidth]{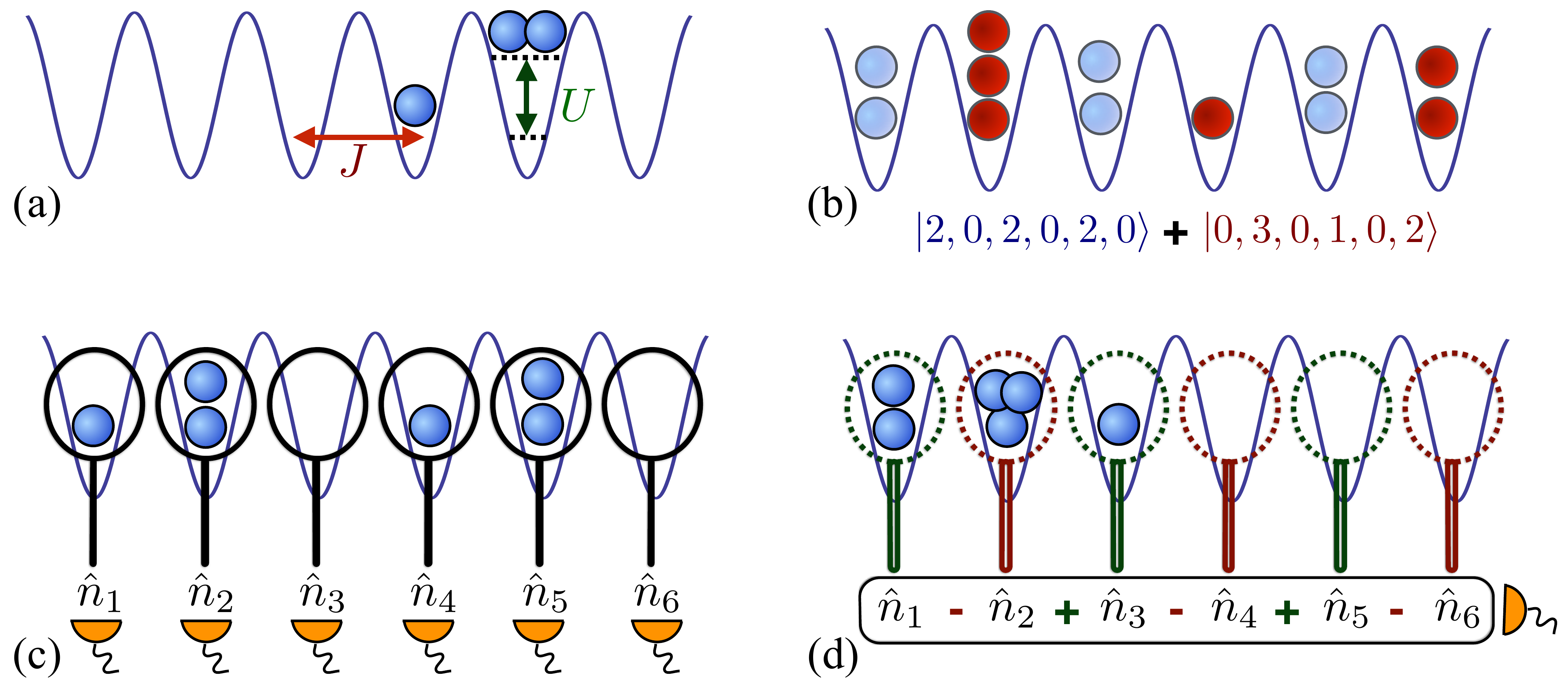}
}
\subfloat[]{
\includegraphics[trim=10 63 15 70,clip,width=\columnwidth]{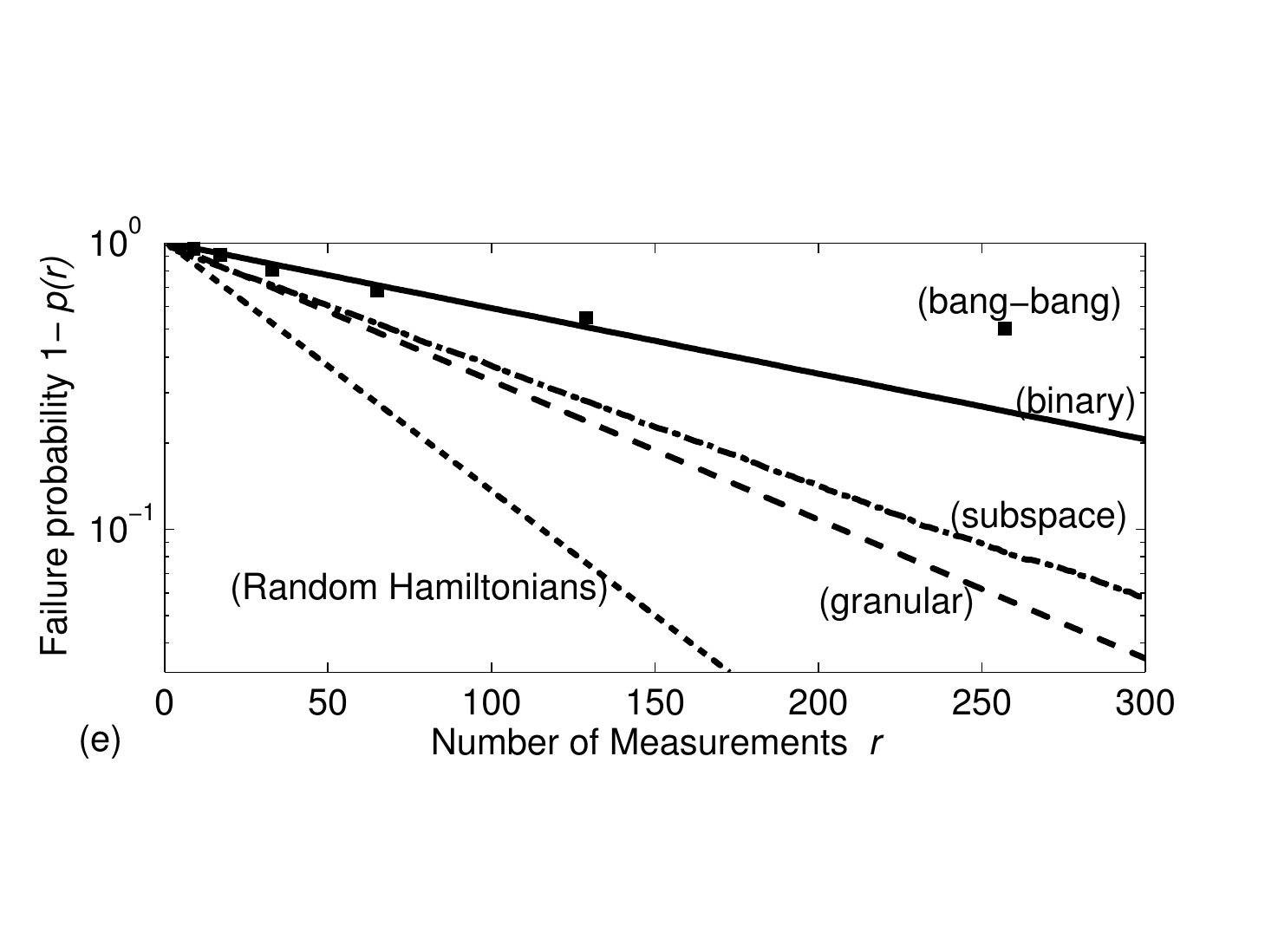}
}
\caption{(color online) (a) The Bose-Hubbard model for ultracold atoms in an optical lattice. (b) Superposition of two multi-mode Fock-states, $\ket{2,0,2,0,2}$ and $\ket{0,3,0,1,0,2}$, each component in a different color. (c) Measurement of a single Fock-state by simultaneous measurement of the atom number in each well. (d) Measurement of atomic number difference between the even and the odd sites. (e) Probability of $r$ consecutive failed measurements for the target state $\ket{\phi_{\text{t}}}=\ket{0,2,0,2,0,2}$. The initial state is the ground state $\ket{\psi_1}$ of the Bose-Hubbard Hamiltonian Eq.~\eqref{eq:BoseHubbard} with $J/U=1.5$. Three different granularities steer the state: Granular measurements (dashed) return a specific Fock-state. Subspace measurements (dot-dashed) yield the distance $D$ [Eq.~(\ref{eq:Distance})] from the target state. Binary measurements (solid) only reveal ``failure'' and ``success''.  For comparison, we show the probability for $r$ consecutive failures for a random Hamiltonian (dotted), and the fidelity of state-preparation via bang-bang control as a function of the number of control cycles $r$ (squares).} 
\label{fig:Example}
\end{figure*}

\emph{Benchmarking --} To prove the general applicability of FUMES outside specific physical systems, we benchmark it against MEDO and MUM for  randomly chosen initial and target states $\ket{\psi_{1}}$ and $\ket{\phi_\text{t}}$, respectively. We will discuss an application to a concrete physical system below; for the moment, we choose the Hamiltonian in an unbiased way by sampling from the Gaussian unitary ensemble, which ensures uniform distribution in the space of Hamiltonians with a particular dynamical time-scale \cite{Zirnbauer1996}. Random Hamiltonians are characteristic for 
 different systems such as chaotic systems of single and many particles \cite{Stockmann2006}. In our case, random Hamiltonians provide a system for which the measurement basis is completely unbiased with respect to the eigenstates of the Hamiltonian. A particular example for such a case is the measurement of a Fock-state in the superfluid regime of the Bose-Hubbard Model, as explained below. 
Our figure of merit is the probability to prepare the target state $\ket{\phi_{\text{t}}}$ after $r$ measurements, $p(r)$ \footnote{In order to rule out the pathological case in which the target state is very close to an eigenstate, we require     $|\braket{\psi_m}{\phi_{\text{t}}}|^2 > 0.1^d$ for all $m$, which amounts to neglecting  $0.3$\% of all Hamiltonians for $d=2$, and even fewer for $d\ge 3$.}.

For MEDO \cite{1742-6596-84-1-012017} [Fig.~\ref{fig:Cartoon}(c)], we assume a set of $r$ operators, each of which projects onto a state
\eq
\ket{\phi_j}=\cos\left(\theta-\frac{\theta j}{r}\right)\ket{\phi_\text{t}}+\sin\left(\theta-\frac{\theta j}{r}\right)\ket{\phi_\text{t}^\perp}, \label{medostate}
\en
where $\ket{\phi_\text{t}^\perp}$ is the  component of $\ket{\psi_1}$ orthogonal to $\ket{\phi_\text{t}}$, $\theta$ is the angle between the initial and the target state with $\ket{\psi_1}=\cos\left(\theta\right)\ket{\phi_\text{t}}+\sin\left(\theta\right)\ket{\phi_\text{t}^\perp}$ and $j\in \{1,...,r\}$. 
In order for MEDO to be successful, all projection operators $\ket{\phi_j}\bra{\phi_j}$ need to be applied subsequently from $j=1$ to $j=r$ and all operators must yield the desired successful outcome. Thus, the success probability of MEDO is the probability that  all measurements be successful,
\eq
p_\text{MEDO}(r)=\cos\left(\frac{\theta}{r}\right)^{2r}  ,
\en
which is independent of the system dimension $d$ \footnote{If the observables are not necessarily equally distributed (MNEDO), the success probability becomes $p_\text{MNEDO}(r)=\prod_{j=1}^r\cos(v_j)^2$ with $\sum_{j=1}^r v_j=\theta$. In fact, $p_\text{MNEDO}(r)$ is  maximized for $v_j=\theta/r$ for all $j$s, i.e.~precisely for the MEDO protocol. This can be shown by induction over the number of measurements $r$ starting from $r=2$ \cite{1742-6596-84-1-012017}, or by contradiction. }. 
 We assume that the measurement operators in MEDO are fixed, such that the protocol does not contain any mechanism to exploit unsuccessful measurement outcomes.

The two maximally unbiased projectors used in MUM \cite{PhysRevA.73.012322} [Fig.~\ref{fig:Cartoon}(d)] are $\hat P_{\text{t}}= \ket{\phi_{\text{t}}} \bra{\phi_{\text{t}}}$ and $\hat P^U=\sum_{j=1}^d \ket{\phi_{\text{t},j}^U} \bra{\phi_{\text{t},t}^U}$, where $|\braket{\phi_{\text{t},j}^{U}}{\phi_{\text{t},j}}|^2=1/d$ for all $j$. That is, each failed measurement of $\hat P_{\text{t}}$ is followed by a measurement of $\hat P_{\text{t}}^U$, which makes a  subsequent successful outcome of $\hat{P}_{\text{t}}$ occur with probability $1/d$. The success probability after $r$ measurements becomes
\eq
p_\text{MUM}(r,d)=1-\left(1-\frac{1}{d}\right)^r, \label{eq:pMUM}
\en
i.e.~the probability of $r$ consecutive failures $1-p(r)$ decays exponentially on a scale defined by $d$. 

The probability of $r$ consecutive failures  $1-p(r)$, i.e.~the probability of not yielding the desired target state after $r$ steps, is shown for MEDO, MUM and FUMES in Fig.~\ref{fig:RandHam}(a). 
The two latter clearly show exponential scaling, and  FUMES consistently outperforms MUM: To understand why, consider a random time-evolution after which the projector $\hat P_{\text{t}}$ is applied  \cite{burgarthDiss}. Such a procedure yields, on average, a target-state occupation probability $1/d$, just as for MUM.  For FUMES, however, each state $\ket{\eta_j}$ characterised by a failed measurement is accompanied by an optimal waiting time that maximises the probability $p_j$ to populate the target state, and, on average,   $p_j > 1/d$. Since the target state can be populated after each measurement with finite probability, $1-p_{\text{FUMES}}(r)$ features an exponential decay, which allows us to define the  \emph{effective dimensionality}  $d_\text{eff}$ via 
\eq 
 p_\text{FUMES}(r) = p_{\text{MUM}}(r, d_{\text{eff}}) ,   \label{eq:effectiveDim} 
\en
which quantifies the performance of FUMES with respect to MUM: For a system of dimension $d$, FUMES requires as many measurements as MUM does for the effective (and smaller) dimension $d_{\text{eff}}$.  The effective dimension $d_{\text{eff}}/d$ decreases with the system dimension $d$ [see Fig.~\ref{fig:RandHam}(b)], because optimising in a higher-dimensional space is more likely to yield a target-state occupation probability $p_j \gg 1/d$. In the unoptimized case with randomly chosen projection times, the effective dimension becomes $d_\text{eff}/d$=1, thus reducing the performance to that of MUM. In other words, even in the presence of substantial errors in the timing of the projection, FUMES does not perform worse than MUM.

Thanks to the $r$ projectors onto states of the form (\ref{medostate}), the performance of MEDO is independent of the system dimension. As illustrated in Fig.~\ref{fig:RandHam}(a), the exponential character of FUMES guarantees that it  always outperforms MEDO if the target failure probability is below a certain threshold. For large system dimensions $d>100$, the target failure probability needs to be below $10^{-2}$ before FUMES becomes favourable, for which, however, the actual feasibility of MEDO becomes improbable.

\emph{Fock-state generation --} We aim at engineering quantum states of multi-mode systems of ultra-cold bosons described by the Bose-Hubbard Hamiltonian \cite{Jaksch1998}
\eq
\hat{H}= -J\sum_{j=1}^{L-1}\left(\hat{a}^\dag_j\hat{a}_{j+1}+\hat{a}^\dag_{j+1}\hat{a}_{j}\right)+\frac{U}{2}\sum_{j=1}^L \hat{n}_{j}(\hat{n}_{j}-1), \label{eq:BoseHubbard}
\en 
where $J$ is the inter-well tunnelling and $U$ is the collisional interaction strength [Fig.~\ref{fig:Example}(a)]. The standard strategies of unitary control  \cite{PhysRevLett.73.58,Viola1998}  or control-free engineering \cite{1367-2630-12-4-043005,Blok2014,PhysRevA.74.052102,Ashhab2010,PhysRevA.86.062331,1742-6596-84-1-012017} require modifications in this setting:  Changing the Hamiltonian parameters abruptly, as for bang-bang control or the SU(2)-decomposition, compromises the lowest-band approximation due to vibrational excitations \cite{Tichy2013}. Continuous changes  impose a speed-limit on achievable operations \cite{PhysRevLett.106.190501}. On the other hand, implementing non-destructive measurements of several precisely chosen non-commuting observables \cite{Ashhab2010,1367-2630-12-4-043005,Blok2014,PhysRevA.73.012322,PhysRevA.74.052102,1742-6596-84-1-012017}  in one experimental realisation remains impractical. 

As an initial demonstration of control, we assume to be equipped with quantum non-demolition measurements of the local atom-numbers \cite{Hume2013,Eckert2007,Mekhov2007,Mekhov2009a,DeVega2008}, which leads to state-collapse onto a Fock-state [Fig.~\ref{fig:Example}(c)] or onto a superposition of Fock-states with certain properties [Fig.~\ref{fig:Example}(b,d)]. For non-vanishing tunnelling $J/U > 0$ and $N>0$, Fock-states are not eigenstates of the Hamiltonian Eq.~\eqref{eq:BoseHubbard}, and they can be target states for FUMES. 
The number of bosons $N$ and lattice sites $L$ determine the system dimension
$d={N +L -1 \choose N } $; here  we consider $N=L=6$, which yields $d=462$.

While a Hamiltonian taken from the Gaussian unitary ensemble is structureless, a natural hierarchy of Fock-states emerges for a system governed by Eq.~\eqref{eq:BoseHubbard} via the distance between two states $\ket{\vec{n}} \equiv \ket{n_1, \dots, n_L}$ and $\ket{\vec{m}}\equiv\ket{m_1, \dots m_L}, $ 
\eq
 D=\sum_{k=1}^L \left\vert\sum_{l=1}^k n_l-m_l\right\vert, \label{eq:Distance} 
\en
which counts the number of tunnelling events required to obtain $\ket{\vec{n}}$ starting from $\ket{\vec{m}}$. The distance $D$ motivates an intermediate level of granularity $M=D$ between binary  ($M=1$) and granular  ($M=d-K$) measurements, which we refer to as \emph{subspace} measurements.

In anticipation of the experimentally relevant infrastructure discussed below, we  illustrate state engineering by FUMES for the target  $\ket{\phi_{\text{t}}}=\ket{0,2,0,2,0,2}$. The probability $1-p(r)$ to remain unsuccessful after $r$ measurements is shown in Fig.~\ref{fig:Example}(e) for $J/U=1.5$, for the three levels of granularity (binary: solid, subspace: dot-dashed, and granular: dashed). 
For all three granularities, numerical optimisation yields an average time between measurements of $2/J$, which corresponds to 0.076~ms for $^{87}$Rb in a conventional 512~nm optical lattice \cite{greinerDiss}.  A finer granularity facilitates faster state-engineering, since it gives more detailed information about the current state of the system, which helps  to choose the optimal time to apply $\hat P_{\text{t}}$. Cumulating a success rate of 99\% with granular measurements thus takes approximately 30~ms. 
 
To set the results into context, we also consider a random Hamiltonian with $d=462$ and fully granular measurements (dotted). The structure of the Bose-Hubbard Hamiltonian implies that states with large distance $D$ need many tunnelling events to be connected, while random Hamiltonians typically couple every pair of states. Therefore, FUMES performs better for random Hamiltonians than for the Bose-Hubbard Hamiltonian.

We also implement bang-bang control \cite{Viola1998}, alternating the Hamiltonians $\hat H_1= \hat H(J/U=0)$ and $ \hat H_2= \hat H(J/U=100)$ (squares). We count the number $r$ of cycles $e^{-i \hat H_2 t_2} e^{-i \hat H_1 t_1}$ and interpret the fidelity of the target state preparation as the success probability $p(r)$. 
 While the non-commutativity of the two Hamiltonians guarantees controllability, bang-bang control \cite{Viola1998} requires more control cycles than measurement cycles needed for FUMES: A measurement in  FUMES strongly perturbs the system, which facilitates the fast population of the target state. In practice, a bang-bang control sequence needs to be chosen in advance, i.e.~before the actual start of the procedure, while   FUMES allows just-in-time optimisation: Every time a measurement fails to produce the target state, the timing for the next attempt is optimised.

\emph{Schr\"{o}dinger cat generation --} As a prominent application of FUMES, we demonstrate the near-deterministic generation of Schr\"{o}dinger-cat states in the Bose-Hubbard model by the experimentally feasible  measurements of the atomic number difference [Fig.~\ref{fig:Example}(d)] between the even and odd sites \cite{Mekhov2007,Mekhov2009a}, 
$ \hat{Z}=\left\vert\sum_{j=1}^L (-1)^j \hat{n}_j \right\vert/2, $
where $\hat n_j$ counts the atoms in site $j$. For vanishing interaction $U \rightarrow 0$, macroscopic superpositions can be generated probabilistically \cite{Mekhov2009a},  conditioned on a non-vanishing measurement result of $\hat Z$. However, this approach typically yields states of very low \emph{macroscopicity} \cite{Frowis2012}, as quantified by 
\eq
 N_\text{eff}(\ket{\psi})=\frac{\max_{\vec{w}}(F(\vec{w},\ket{\psi}))}{4L} ,
\en
where $ F(\vec{w},\ket{\psi}) $ is the   quantum Fisher-information
$ F(\vec{w},\ket{\psi})=4(\langle\psi|\hat{S}(\vec{w})^2|\psi\rangle-(\langle\psi|\hat{S}(\vec{w})|\psi\rangle)^2), $ 
and $\hat{S}(\vec{w})$ is the measurement operator
$ \hat{S}(\vec{w})=\sum_{j=1}^L w_j \hat{n}_j=\vec{w}\cdot\vec{\hat{n}}, $ 
where we restrict $\vec{w}$ to $\{\pm 1\}^L$, accounting the impossibility to directly measure superpositions of different particle numbers without auxiliary reservoirs \cite{RevModPhys.79.555}.  The macroscopicity $N_\text{eff}(\ket{\psi})$ is the  minimal number of particles for which the validity of quantum mechanics is required for a description of the observed macroscopic fluctuations \cite{Frowis2012}. A Schr\"{o}dinger-cat state fulfils $N_\text{eff}=N$, the ground-state of the Bose-Hubbard Hamiltonian for $J/U=1.5$ and $d=N=6$ gives $N_\text{eff}=0.87 \approx N/7$, while single Fock-states do not carry any macroscopic entanglement and yield $N_{\text{eff}}=0$.

\begin{table}[htbp]
\caption{Macroscopicity $N_{\text{eff}}$ of  states sampled from the full Hilbert space and subspaces with specified $Z$ for $N=d=6$.}
\label{tab::Means}
\begin{tabular}{c | c c c c c}
Space & Hilbert& $Z=0$ & $Z=1$ & $Z=2$ & $Z=3$\\
\hline
$\langle N_\text{eff}(\psi) \rangle$ & 1.83(6) & 1.6(1) & 1.65(9) & 2.64(3) & 5.9(1) \\ 
\end{tabular}
\end{table}

\begin{figure}[tbp]
\includegraphics[trim=11 75 14 80,clip,width=\linewidth]{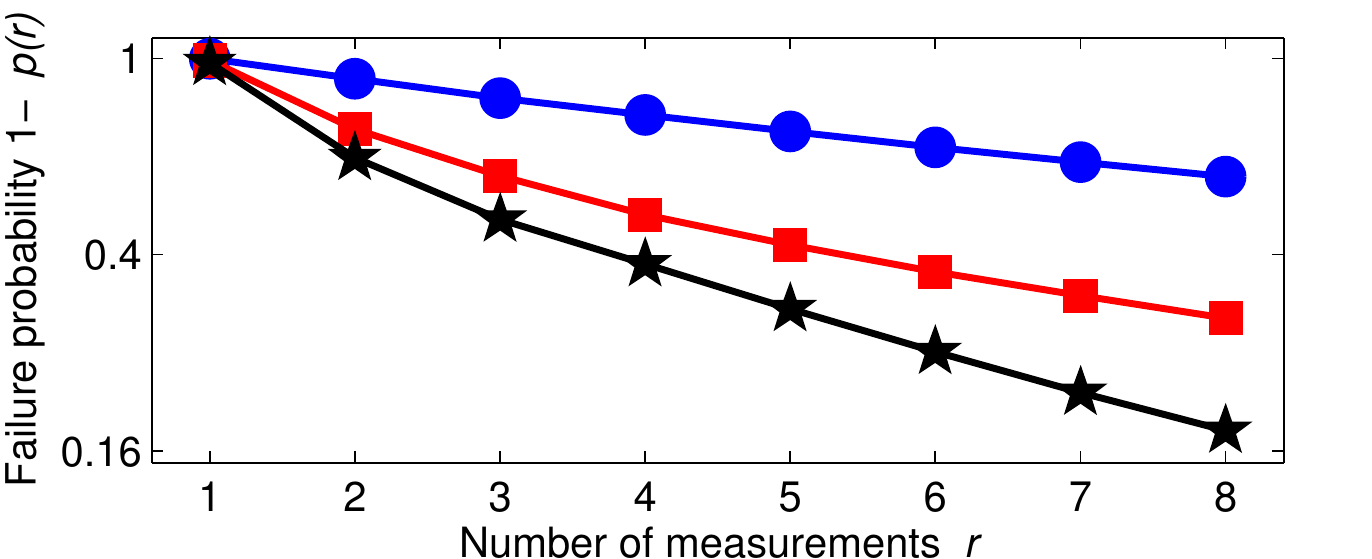}
\caption{(color online) Probability not to yield a state in the Schr\"odinger cat target-subspace $Z=3$ after $r$ measurements with FUMES, for  $J/U$=0.25 (blue circles), $J/U$=0.50 (red squares), $J/U$=1.50 (black stars). }
\label{fig::NumMes}
\end{figure}

The macroscopicity within subspaces of constant $Z$ has a small spread and is given in Table.~\ref{tab::Means}. Thus, finding $Z=N/2$ is necessary to achieve Schr\"odinger-cat-like macroscopicity characterised by $N_{\text{eff}} \approx N$.   However, measuring $Z=N/2$ in the ground state of the Bose-Hubbard model is highly unlikely: For a superfluid state of 100 bosons in 100 sites, the probability is less than $10^{-30}$, and even for $N=6$ particles and $L=6$ sites with $J/U=1.5$, the probability remains less than 2\%. 

Using FUMES, we can exploit the tunnelling dynamics induced by the Hamiltonian to steer the initial state into the otherwise improbable subspace $Z=N/2$ [Fig.~\ref{fig::NumMes}]. 
The target space $Z=N/2=3$ has dimension 56, which facilitates state engineering in comparison to Fock-state generation [Fig.\ref{fig:Example}~(e)]. Larger values of $J/U$ come with super-fluid-like eigenstates that mix more Fock-states than low values leading to Mott-insulator-like eigenstates. This makes larger values of $J/U$ more favourable, consistent with Fig~\ref{fig::NumMes}. Eight measurements then suffice to accumulate a success rate of more than 80\% for $J/U=1.5$, while the typical time between two measurements for $J/U=1.5$ remains around $2/J$, as for the Fock-state target.

\emph{Conclusion --} The minimal requirements for unitary control are the availability of two non-commuting Hamiltonians $\hat H_1$ and $\hat H_2$ \cite{Viola1998}. We translated this  setup into the realm of control-free engineering: A fixed Hamiltonian $\hat H$ and a projector onto the desired target subspace $\hat P_{\text{t}}$ then permit FUMES as long as the dynamics induced by the Hamiltonian effectively steers quantum states into the target subspace. In situations in which the Hamiltonian is uncontrollable or no tailored measurements can be applied, FUMES emerges as a natural way to perform state engineering. Its computational costs are manageable even for large system dimensions, since the waiting time until the next measurement can be optimised in a just-in-time fashion. Thanks to the optimisation of the waiting time, it outperforms methods for which the success probability of the measurement scales as $1/d$ \cite{PhysRevA.73.012322,burgarthDiss}. Timing errors may affect the overall success probability of FUMES, since inaccuracies in timing will result in projections at non-optimal moments in time and, consequently, to a lower probability to reach the target. However, timing errors will not affect the actual fidelity of the preparation, which is entirely defined by the fidelity of the projection operator. 

Our greedy  strategy that always aims at the desired target in the next measurement  may be refined further: For granular measurements, the trajectory of a state steered by  FUMES corresponds to a classical random walk on $d-K$ nodes: $d-K-1$ nodes correspond to non-target states, one represents the target subspace. The probability to jump from one node to the other is then optimised to achieve the largest probability to populate the target node. For a state with very low -- albeit optimised -- success probability, it can be advisable to induce a detour and rather aim at another non-target state, provided it comes with a large probability to reach the target in a subsequent measurement.  Alternatively, it is worthwhile to combine FUMES and MEDO, i.e.~to optimize, both, the time-evolution and the form of the next projector to be, under the given experimental limitations. Such hybrid protocol would likely saturate the general possibilities for control, exploiting all available resources. Besides the assessment of such more sophisticated strategies, it remains to be studied which ensembles of states and Hamiltonians can be engineered, and how the strategy performs on systems beyond the Bose-Hubbard-Hamiltonian.

\emph{Acknowledgements} 
The authors would like to thank S{\o}ren Bendlin Gammelmark for contributions to the numerical code and illuminating discussions, to Minsu Kang for clarifying remarks on macroscopicity, and to Klaus M{\o}lmer for very helpful comments on the manuscript. Financial support from the Lundbeck Foundation and the Danish National Research Council is gratefully acknowledged.


\end{document}